\documentclass[namedreferences]{solarphysics}

\usepackage[hyperref,optionalrh,showbiblabels]{spr-sola-addons} % For Solar Physics 
\usepackage{graphicx}        % For eps figures, newer & more powerful
\usepackage{color}           % For color text: \color command
\usepackage{breakurl}        % For breaking URLs easily trough lines
\usepackage[normalem]{ulem}            % \sout{} for strikethrough edits
\newcommand{\ie}{{\it i.e.},}

\newcommand{\degree}{\ensuremath{^\circ}}

\newcommand{\aap}{    {\it Astron. Astrophys.}}

\newcommand{\aj}{     {\it Astron. J.}} 
\newcommand{\apj}{    {\it Astrophys. J.}}
\newcommand{\apjl}{   {\it Astrophys. J. Lett.}}

\newcommand{\solphys}{{\it Solar Phys.}}

\chardef\us=`\_

\begin{document}

\begin{article}
\begin{opening}

\title{Deciphering Solar Magnetic Activity. 
The Solar Cycle Clock }

\runningtitle{The Solar Cycle Clock}
\runningauthor{R.\ J.\ Leamon {\em et~al.}}

\author[addressref={1,2},corref,email={robert.j.leamon@nasa.gov}]%
{\inits{R.J.}\fnm{Robert J.}~\lnm{Leamon}~\orcid{0000-0002-6811-5862}}
\author[addressref={3}]%
{\inits{S.W.}\fnm{Scott~W.}~\lnm{McIntosh}~\orcid{0000-0002-7369-1776}}
\author[addressref={4}]{\inits{A.M.}\fnm{Alan M.}~\lnm{Title}}

  \address[id={1}]{University of Maryland--Baltimore County, Goddard Planetary Heliophysics Institute, Baltimore, MD 21250, USA}
  \address[id={2}]{NASA Goddard Space Flight Center, Code 672, Greenbelt, MD 20771, USA.}
  \address[id={3}]{National Center for Atmospheric Research, P.O. Box 3000, Boulder, CO~80307, USA}
  \address[id={4}]{Lockheed Martin Advanced Technology Center, 3251 Hanover Street, Building 252, Palo Alto, Colorado 94304, USA}

\begin{abstract}
The Sun's variability is controlled by the progression and interaction of the magnetized systems that form the 22-year magnetic activity cycle (the ``Hale Cycle'') as they march from their origin at $\sim$55\degree\ latitude to the equator, over $\sim$19 years. 
We will discuss the end point of that progression, dubbed ``terminator'' events, and our means of diagnosing them. Based on the terminations of Hale Magnetic Cycles, we construct a new solar activity `clock' which maps all solar magnetic activity onto a single normalized epoch. The Terminators appear at phase $0 * 2\pi$ on this clock (by definition), then solar polar field reversals commence at $\sim${}$0.2 * 2\pi$, and the geomagnetically quiet intervals centered around solar minimum, start at $\sim${}$0.6 * 2\pi$ and end at the terminator, lasting 40\% of the normalized cycle length. With this onset of quiescence, dubbed a ``pre-terminator,'' the Sun shows a radical reduction in active region complexity and (like the terminator events) is associated with the time when the solar radio flux crosses F10.7=90\,sfu---effectively marking the commencement of solar minimum conditions. 
In this paper we use the terminator-based clock to illustrate a range of phenomena 
that further emphasize the strong interaction of the global-scale magnetic systems of the Hale Cycle.

\end{abstract}
\end{opening}

%-------------------------------------------------

\section{Introduction}\label{sec:intro}
Sunspots have been considered the canon of solar variability since \cite{1844AN.....21..233S} first noticed the 11-year or so oscillation in their number. Schwabe inspired Wolf to make his own daily sunspot observations \citep{1861MNRAS..21...77W} and then extend the record back another 100 years using the earlier observations of Staudacher 1749 to 1787, Flaugergues from 1788 to 1825, before Schwabe's 1826 to 1847 record \cite[e.g.,][]{Arlt2008,2015LRSP...12....4H}. Wolf then numbered the (complete) cycles from his first minimum in 1755--56; the Cycle~24 minimum was recently announced as occurring in December~2019 and we have already witnessed the first activity of Cycle~25. We know the recent sunspots ``belong'' to Cycle~25 because the leading polarity spot of each cycle alternates, and at approximately the maximum of each sunspot cycle, the orientation of the Sun's dipole magnetic field flips. The more fundamental period of solar activity is thus the 22-year magnetic cycle, or ``Hale cycle'' after \cite{Hale19}.

In the century since Hale, much effort has been put into modelling the solar cycle \citep{Bab61,Lei69} and forecasting the number of spots, culminating in official NASA and NOAA prediction panels \citep{1997EOSTr..78..205J,Pesnell2008,2019AGUFMSH13B..03B}. 
% [Joselyn (23); Pesnell (24); Biesecker & Upton (25)]. 
These reports were issued at the minimum of their respective cycles.

Recently, we introduced the concept of the 
``Terminator,'' 
as a new, physically-motivated, means of timing the onset of solar cycles \citep{McIntosh2019,LeamonEA2020}. The canonical measure, the (committee-defined) minimum of the sunspot number, is physically arbitrary 
% [REFs: STEVE, AMPLITUDE]. 
\citep{2020arXiv201006048M,McIntoshEA20b}.
Sunspot minimum is defined during an epoch where the cumulative effect of four simultaneously decreasing and increasing quantities---the number of new and old cycle polarity spots in each hemisphere as the magnetic systems on which they exist---(significantly) overlap in time. 
Considering a precise date---identically when there is no more old cycle polarity flux left on the disk---a terminator marks the end of a Hale magnetic cycle. Terminators are inextricably linked to the evolution of the ``Extended Solar Cycle'' \citep{1983A&A...120L...1L}; as a proxy of the Hale Cycle, and illustrate that there is far more complexity to the Sun's magnetism than sunspots---the global scale interaction of the Hale Cycle's magnetic systems shape the output of our star and its evolution. We are just beginning to explore the expressions of this behavior.

\subsection{The Extended Solar Cycle}

{% \color{blue}
Our companion paper \citep{2020arXiv201006048M} covers the history of our understanding of the Extended Solar Cycle up to the late 1980s/ early 1990s \citep{Wil88,Har92,1994ssac.book.....W}, and our recent efforts to revise and extend the concept,
by comprehensively studying the {\em observed\/} patterns of magnetic solar activity from 
filaments, % (hence the `140 Years' in the title), 
coronal emission lines,
the rush to the poles,
and the torsional oscillation,
in addition to the counting of spots.}
Here we expand on and build on the ESC framework of Wilson and our preceding works, including the concept of the Terminator, to explain the 
Activity
Cycle as a clock, albeit one whose hour hand moves at a different rate every 
{revolution, 
but at a consistent rate within that revolution (\ie{} 11-or-so-year activity cycle),
{% \color{blue}
and whose landmarks are explained by the actions and interactions of the overarching, overlapping, 22-year Hale Magnetic Cycles.}

The title of this paper reflects our views on the regularity of solar activity.
We do not concern ourselves here with the dynamics of the deep solar interior---whether 
\citep[{\em e.g.},][]{Dicke78,2019RvGeo..57.1129R} or not
\citep[{\em e.g.},][]{Hoyng96} there is a regular deep-seated ``chronometer'' at or below the solar tachocline, or too greatly with the dynamo itself.
Rather, 
as we shall see,
the observed magnetic, radiative, and eruptive activity that arises from our star's dynamo action is coherent within each activity cycle, and results from the overlap of two Extended Solar Cycles 
(Hale Cycles).

}

\subsection{Coronal Bright Points}
\label{sec:cbp}

\begin{figure}[p]
\centering
\includegraphics[width=0.9\linewidth]{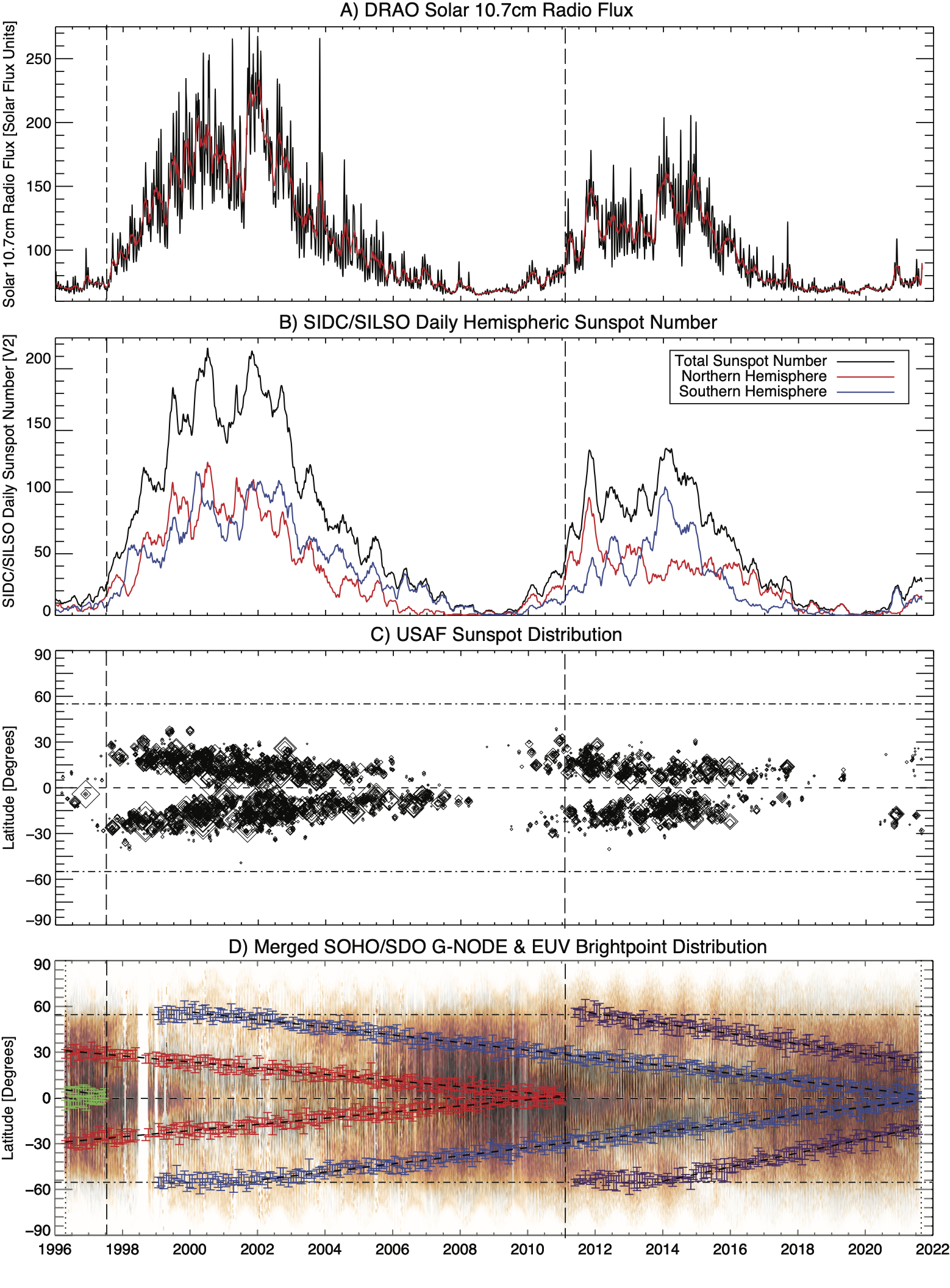}
\caption{Updated and  extended version of Fig.~1 of \cite{LeamonEA2020}, demonstrating the concept of Terminators and the brightpoint-activity band model and their relevance to the sunspot number during the {\em SOHO\/} epoch (January 1996--August 2021). Panel A shows the daily (grey) and 50-day smoothed (red) 10.7cm solar radio flux from the Dominion Radio Astrophysical Observatory (DRAO). Panel B shows the daily (v2) hemispheric (red-north, blue-south) and total (black) sunspot numbers from the Solar Influences Data Center (SIDC) of the Royal Observatory of Belgium, Brussels. Each of the sunspot time series has a running 50-day smoothing. Panel C shows the United States Air Force (USAF) sunspot record---the size of the diamonds reflects the relative area of the sunspots in the record. Panel D shows the tracked centroids of the BP distribution for each hemispheric activity band, extending the work of \cite{Mac14}---cycle 22 bands in green, cycle 23 in red, 24 in blue and 25 in purple. The dashed vertical lines indicate the terminators \citep{McIntosh2019} of cycles~22 (August 1997) and~23 (February 2011). 
Extrapolating the fit for 
cycle 25 (straight line through purple symbols) predicts a termination date of October $2031 \pm 9$ months.
}
\label{fig:bands}
\end{figure}

Our preceding work on the (extended) solar cycle
\citep{Mac14,McIntosh2019,LeamonEA2020,2020arXiv201006048M} has been driven through the study of 
ubiquitous small features observed in the Sun's extreme-ultraviolet corona, ``EUV Bright Points,'' or BPs \citep{1974ApJ...189L..93G, 2003ApJ...589.1062H, 2005SoPh..228..285M}.
What \cite{1973SoPh...32..389H}
%% Harvey
referred to as ``ephemeral active regions'' we now understand to be (the larger) BPs.
The drift behavior of Bright Points relative to solar rotation implies that they are deeply rooted below the surface, and thus are associated with the evolution of the rotationally-driven giant convective scale cells \citep{2014ApJ...784L..32M} that had vertices that were dubbed ``g-nodes.'' 
Together, these features permit the tracking of the magnetic activity bands of the 22-year magnetic cycle of the Sun that extend the conventional picture of decadal-scale solar variability. 
Thus the culmination inference of our (and colleagues') preceding work is that
the global-scale (intra- and extra-hemispheric) interaction of these magnetic activity bands was required to explain the appearance and evolution of sunspots on the magnetic bands and thus to shape the (extended) solar cycle, as illustrated in Figure~\ref{fig:bands}.

Figure~\ref{fig:bands}, 
an extended, updated, 
version of Fig.~1 of \citet{LeamonEA2020}
illustrates the evolution of EUV BPs from 1996, at the minimum between sunspot cycles~22 and~23, to the present 
(through 
July~2021)
in context with the sunspot number, their latitudinal progression, and a signature measure of the Sun's radiative output---the 10.7cm solar radio flux. The large-scale magnetic activity bands that combine to shape sunspot cycles 22, 23, 24, and 25 are identified, as are the terminators. Note that, in both 1997-98 and 2010-11, the sunspot number has already started to increase from its activity minimum nadir since the bands temporally overlap. 
This is readily observed in comparison with panel C. 
Notice also the ``clumps'' of sunspots produced in each hemisphere and their corresponding signature in the total and hemispheric sunspot numbers \citep{2015NatCo...6.6491M, 2017NatAs...1E..86M}.
Further, the terminators are clearly associated with a rapid increase in activity in (at least) one solar hemisphere \citep{Mac14}.
For example, the cycle 23 sunspots did not appear to grow in abundance or size until the cycle 22 bands had terminated (in late 1997). Similarly, the polarity mirror-image of this progression occurred in early 2011 for cycle 24 sunspots, following the termination of the cycle 23 bands. This equatorial termination, or cancellation, appears to signal the end of one sunspot cycle and leaves only the higher-latitude band in each hemisphere. Sunspots rapidly appear and grow on that mid-latitude band for several years in this, the ``ascending phase,'' until the next (oppositely-signed) band appears at high latitude. The presence of the new oppositely signed band triggers a downturn in sunspot production on the mid-latitude band; this occurrence defines the maximum activity level of that band and the start of a new extended cycle.

\cite{2019NatSR...9.2035D} suggested that the most plausible 
mechanism for rapid transport of information 
from the equatorial termination of the old cycle's activity bands 
(of opposite polarity in opposite hemispheres)
to the mid-latitudes to trigger new-cycle growth
was a solar ``tsunami'' in the solar tachocline that migrates poleward with a gravity wave speed 
($\sim$300$\mbox{ km s}^{-1}$).

\subsubsection{Cycle 24?}

At the time of 
original submission
(December 2020), activity ramped up rapidly and the daily F10.7 first exceeded 90~sfu on November~23, 2020, reaching a peak of 115~sfu on November~29, 2020, before falling back.
On the 26-year scale of Figure~\ref{fig:bands}, the SSN (Panel B) and distribution of sunspots (Panel C) perhaps 
do not demonstrate 
this
surge in activity.
Some, but not all measures of activity 
met the thresholds for declaring the Cycle~24 terminator \citep{Mac14,McIntoshEA20b}.
After falling back to a minimum value 
below 70~sfu 
in January~2021, F10.7 rose slowly, but steadily into the 90s through June~2021 and then surged to 100~sfu into July 
{% \color{blue}
before again falling back to a minimum value 
below 80~sfu in August.
}
It is unfortunate that this paper could not be published before the first X-flares of 
Cycle~25: the X1.5 event at 14:29UT on July~3, 2021
{% \color{blue} 
and the X1.0 event at 15:17UT on October 28, 2021},
but that does not affect our ability in what follows to describe and discuss historical data.

\subsection{Terrestrial Effects}

The terminators lead to a rapid (within about one solar rotation) surge in all forms of solar output---radiative, particulate and ejecta.
The jump in F10.7 radio emission and sunspots is shown in Figure~\ref{fig:bands};
\cite{2021ESS....801223L}
%% Nina
showed that 
the ratio of the pre- and post-termination emission from SDO/EVE across that temperature range scales from 8\% to 85\% and is highly localized with plasma emission around 5 million Kelvin. 
This behavior has also been noted, without explanation, by two recent studies \citep{2017ApJ...844..163S,2017SciA....3E2056M}.

% Alan Title
\cite{2015TESS....140602T} discussed the bursty nature of X-flares, that is, they occur in clusters, with shorter ($\simeq$ day) gaps between individual flares, and then months between clusters.
Such clustering timescales are consistent with the period of a Rossby wave propagating in the solar tachocline \citep{2017NatAs...1E..86M}.
% Shuggie
\cite{2020SoPh..295..132H}
has also tried to explain clustering, 
but in terms of a ``waiting time'' relaxation oscillator, but he also notably remarked that there is a ``last best flare'' for each solar cycle \citep{2018TESS....131916H}, 
as has \cite{Chapman2020}.
Similarly, the first X-flare 
(or flare cluster)
of a cycle also seems to happen within a rotation or two of the terminator.
We shall 
explore this phenomenon and 
demonstrate
the rationale for it in much detail below.

\cite{2021ESS....801223L} also introduced the concept of the ``standard cycle'' or unit cycle based on the terminator-terminator phase, 
which we will explore more fully in this paper.
Rather than defining a standard superposed epoch analysis \citep{1913RSPTA.212...75C}
repeating over some number of days, the critical modification is to first scale time to be fractions of a cycle, from terminator to terminator.
(One may think of this then as a ``phase'' of the solar cycle, but we choose here to express time in terms of a fraction 0--1 rather than 0--$2\pi$.)
\cite{LeamonEA2020} put the solar cycle predictability based on the linear extrapolation of BP trajectories
shown in Figure~\ref{fig:bands}(d) 
on a more mathematically rigorous footing, 
based on Hilbert transforms of solar activity indices.
\cite{Chapman2020} extended the Hilbert transform formulation of \cite{LeamonEA2020}, and used it to create a ``solar cycle phase clock'' 
and mapped each of the last 18 solar cycles onto a single normalized 11 year epoch.
\cite{Chapman2020} then found that there was a predictable start and end to significant geomagnetic activity (X-class flares and high $aa$ index: only two $aa>300$~nT geomagnetic storms happened inside the quiet interval in the entire 1868--present record of the index). 
The quiet interval is 40\%\ of the cycle, ends {\em at\/} the terminator, 
and is roughly centered on solar minimum.

Although we led development of the terminator-to-terminator phase clock, 
Owens, Lockwood and collaborators have also 
considered a modified superposed epoch:
\cite{2014SoPh..289..407T}
expressed the solar cycle as a phase (expressed as 0--360\degree) from polar field reversal to polar field reversal to study differences in GCR flux during alternate solar cycles. 
The Hale Cycle is particularly clear in GCR flux; the 
different particle drift patterns when the northern solar pole has predominantly positive (denoted as a $qA>0$ cycle) or negative ($qA<0$) polarities
give rise to distinct 
flat-topped GCR time series for $qA>0$ cycles and more peaked for $qA<0$ \citep[see also][]{2021ESS....801223L}.

More recently (submitted almost simultaneously with this paper), 
\cite{2021SoPh..296...82O}
used a minimum-to-minimum phase (expressed as 0.0--1.0 of a cycle) to 
report on the difference in extreme-event 
(defined as $aa > 300$~nT)
occurrence during odd- and even-numbered solar cycles, with events clustering earlier in even cycles and later in odd cycles. 
They interpreted this finding in terms of the overlying coronal magnetic field---extreme events require a positive polarity in the northern polar cap and thus a North-to-South dipolar solar magnetic field---and the polar field reverses at maximum
(see the cartoon that is their Figure~9).

We argue, though, that as 
Figure~\ref{fig:bands} and \cite{Chapman2020}, \cite{Mac14}, and \cite{McIntosh2019} show, 
(1) the Hale Cycle (terminator) clock is more fundamental than a minimum-to-minimum phase or maximum-to-maximum phase, and
(2) edges are easier to detect and determine precisely than the extremum values of (summed) quantities, such as (hemispheric) sunspot number and the magnetic field strengths at the two poles,
we shall continue here with the Hale Cycle clock.
(Continuing the clock analogy, if the reader wishes to directly compare our figures with those of \cite{2021SoPh..296...82O}, one may think that they ``are in a different time zone to us'' and are approximately 0.2~cycles ahead.)

We have already remarked that
\cite{1981SoPh...70..173L}
% Legrand and Simon
also considered the solar cycle dependence of geomagnetic activity, 
deducing that the 
recurrent streams of the latter part of the sunspot cycle and associated behavior of the $aa$ index were best explained in terms of 
the extended cycle,
that is, due to the simultaneous presence on the disk of both current- and next-cycle magnetic elements.
Specifically, the late-cycle coronal holes that give rise to 27-day recurrent high speed streams are due to what will become the sunspot-generating activity band of the next cycle some 5--7 years later.
Thus,
\cite{1981SoPh...70..173L}
provided a foundation for geomagnetic precursor solar cycle predictor methods \citep[e.g.,][]{Feynman82},
which is the general method that has proved most successful for retroactive and forward cycle prediction \citep{1997EOSTr..78..205J,Pesnell2008}.

Finally, the
composition of the solar wind, such as the elemental abundance of Helium, or Fractionation of Iron has been observed to vary with the solar cycle 
\citep[e.g.,][]{2007ApJ...660..901K,2011ApJ...740L..23M,2012ApJ...745..162K,2013ApJ...768...94L,2019ApJ...879L...6A};
all reflect the physical properties of the corona and its evolution over time,
specifically the (electron) temperature for ionisation and charge equilibrium,
the magnetic topology, and level of (turbulent) electromagnetic fluctuations. 
The fast wind from coronal holes, slow wind, and ejecta all show different behaviours, with time.
In particular, the abundance of Helium as a function of wind speed is relatively constant over the cycle for the fastest wind, but the slowest wind shows a factor of $\sim$5 variation from minimum to maximum, with the biggest changes seemingly coincident with the biggest changes in F10.7 and spectral irradiance
\citep[e.g.,][]{2017SciA....3E2056M,2021ESS....801223L}.

We say ``seemingly,'' 
not as a weasel word, but
because previous
studies have binned the solar wind by velocity, and averaged the Helium abundance or other composition measurements over long ($\sim$250 day) intervals, 
making comparisons with the sudden changes in the corona 
such as at the Terminators 
described above challenging. 
However,
\citet{2021SoPh..296...67A}
% Alterman, Kasper, Leamon and McIntosh, 
% \footnote{``Helium Abundance Heralds the Onset of Solar Cycle 25,'' submitted to {\em Solar Physics\/} and ArXiv.}
%% delete footnote because it's now out!
do suggest that the abrupt depletion of Helium Abundance to its cyclic minimum that occurred in late 2018--early 2019 was due to the time difference in new cycle BPs appearing in one hemisphere before the other affecting the global magnetic connectivity and level of Alfv\'enic fluctuations, thus
gving an affirmative answer to
% Kasper
\cite{2007ApJ...660..901K}, who
asked
``Is it possible that there is an insufficient density of Alfv\'enic fluctuations to impart a differential flow in the corona where very slow wind emerges during solar minimum?''

\subsection{Outline}

Based on \cite{Chapman2020} and \cite{LeamonEA2020}, we 
extend their work and 
investigate the effects of the Terminator on the global solar magnetic field, active region complexity, X-flare production, and coronal emission. 
We will demonstrate that bands model of \cite{Mac14} can explain the occurrence of not just the terminators but also a ``pre-Terminator,'' 
on the decline phase,
in a ``clocked'' unit solar cycle, 
replacing the 11-ish year cycle by a ``solar cycle activity phase,''
{% \color{blue}
all explained by landmarks of the 22-year Hale Magnetic Cycle.}
The pre-Terminator \citep[named by][]{Chapman2020} has as similar dramatic effects as the Terminator
that has on the ramp up in solar activity. 
We will further show that the start and end of the quiet interval, 
bookended by the pre-Terminator and Terminator, with minimal geo-effective/ socio-economic space weather impacts, 
are eminently forecastable.

\section{Data Sources and Selection}
\label{sec:data}

This study uses a wide range of solar remote observations, from both space-borne satellites and ground-based observatories.

From space,
the primary motivator of all activity bands studies is the determination and tracking of EUV BPs from
The \emph{Solar and Heliospheric Observatory} (SOHO) \emph{Extreme-ultraviolet Imaging Telescope} (EIT) \citep{1995SoPh..162..291D} and \emph{Solar Dynamics Observatory} (SDO) \emph{Atmsopheric Imaging Assembly} (AIA) \citep{2012SoPh..275...17L} telescopes (in 195\AA\ and 193\AA, respectively).
Following \citet{Mac14}, we identify BPs in a manner that accounts for differences in the two instruments.
SDO also provides spectral irradiance data via the \emph{EUV Variability Experiment} (EVE) \citep{2012SoPh..275..115W}.
The other space-based dataset used is the
Geostationary Operational Environmental Satellites
(GOES) X-ray flare catalog%
\footnote{{\tt www.ngdc.noaa.gov/stp/space-weather/solar-data/solar-features/solar-flares/ x-rays/goes/xrs/}} combining measurements from the series of 17 GOES satellites from 1976--present \citep{2009SPIE.7438E..02C}.

From the ground, the 
Wilcox Solar Observatory\footnote{{\tt wso.stanford.edu/HCS.html}} \citep{1977SoPh...54..353S} provides solar magnetic field measurements 
(also 1976--present).
We also use 
longer-term synoptic ground-based solar observations, including the 
Penticton 10.7cm radio flux\footnote{{\tt spaceweather.gc.ca/solarflux/sx-4-en.php}}  \citep{2013SpWea..11..394T}; 
and the
NGDC composite Coronal Green Line (Fe~{\sc xiv}) data\footnote{{\tt www.ngdc.noaa.gov/stp/solar/corona.html}} \citep{1994SoPh..152..153R}.
Sunspot number data is from 
the Solar Information Data Center \citep[SIDC;][]{SIDC,Vanlommel2005}, 
and sunspot {\em area\/} from the combined Royal Greenwich Observatory/ United States Air Force record, originally compiled and maintained by NASA's Marshall Space Flight Center\footnote{{\tt solarscience.msfc.nasa.gov/greenwch.shtml}} 
\citep{2015LRSP...12....4H}.

\section{Observations}
\label{sec:results} 

\subsection{Solar Magnetic Fields}

We first turn our attention to the large-scale solar magnetic fields, as measured by Stanford's Wilcox Solar Observatory. 
As a matter of course, they produce synoptic measurements of the sun's magnetic field, as observed at the photosphere, 
the coronal magnetic field as calculated from photospheric field observations using various potential field models, and
the tilt angle of the Heliospheric Current Sheet as deduced from the coronal extrapolations. 
We shall focus here on the polar field strength, and the 
lower
moments 
of the multipole field expansion \citep{1969SoPh....6..442S,1977SoPh...54..353S}.

\begin{figure}[p]
\centering
\includegraphics[angle=90,width=\linewidth]{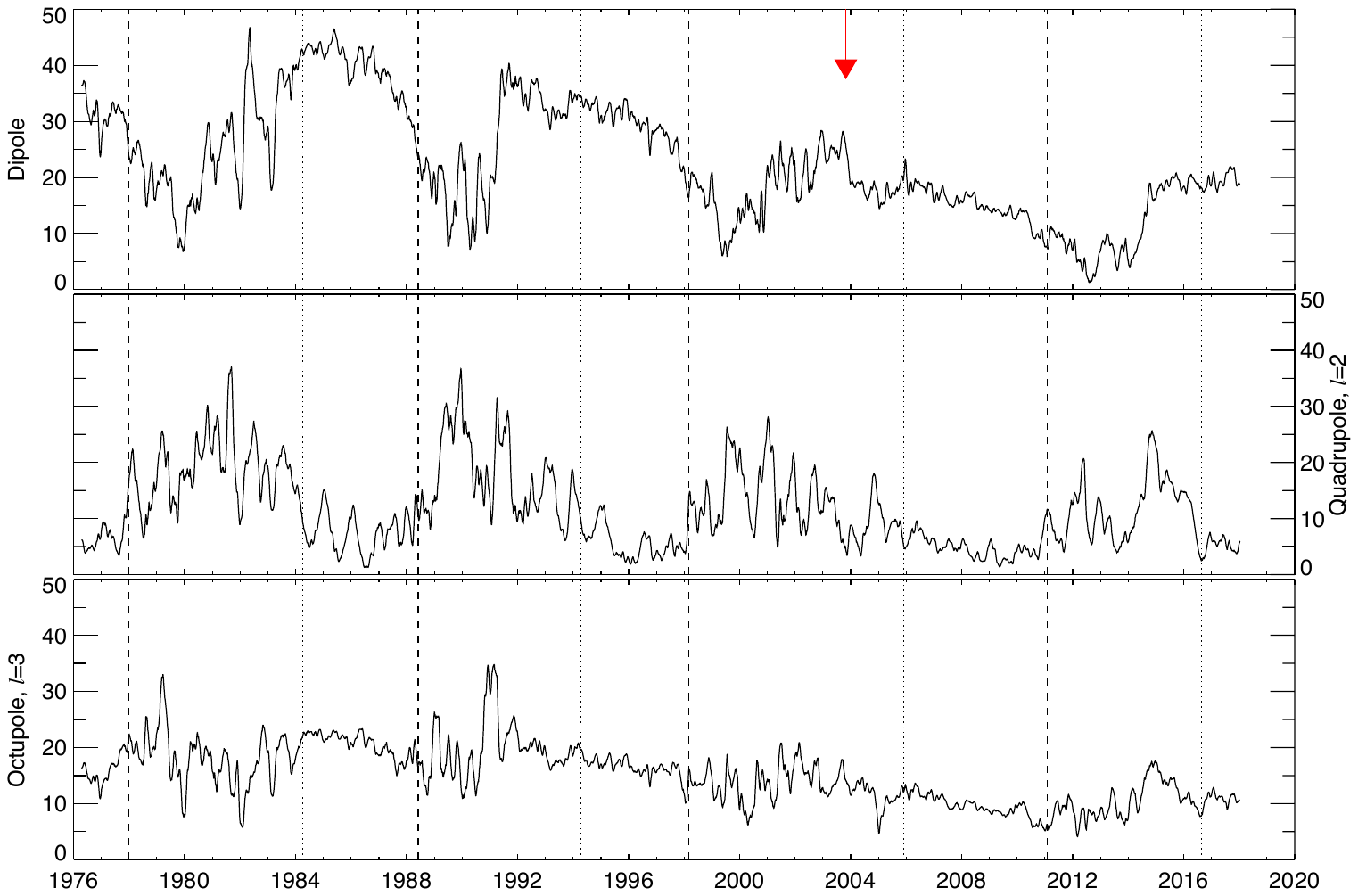}
\caption{Evolution of the solar dipole and multipole components from the Wilcox Solar Observatory expansion of the coronal magnetic field. Dashed lines represent terminators; dotted lines represent the pre-terminators. The red arrow corresponds to the Halloween Storms of 2003---note sharp drop in dipole field strength (top panel) with no parallel in any other cycle.}
\label{fig:wt}
\end{figure}

Figure~\ref{fig:wt} shows the time history of the dipole, quadrupole and octupole moments over the whole Wilcox record.
The 11-ish year cyclic nature of the dipole moment is clear (and not at all surprising).
The dashed lines represent the Terminators, as defined by \cite{McIntosh2019} and \cite{LeamonEA2020}. 
As solar activity jumps up at the terminators, we see the dipolar moment drop, on its way to a minimum at the solar maximum polar field reversal, and an uptick in the quadrupole moment. 
The solar dynamo does not really suddenly gain a quadrupole or higher multipole moment with the onset of activity; rather one should think of the higher orders as (averaged) interpretations of the complexity of the active region magnetic fields.

There are two interesting features in Figure~\ref{fig:wt},
beyond the expected cycling behavior.
The first, 
singular, interesting feature
is marked by the red arrow, and
which corresponds to the ``Halloween Storms'' of October--November 2003.
\cite{LeamonEA2020}
mooted the idea that those eruptions---the largest of Cycle~23, and possibly/ probably larger than the Carrington event 
% [REFS -- Carrington, Tsurutani, Baker]
\citep{Carrington1859,Hodgson1859,2003JGRA..108.1268T,2013SpWea..11..585B}---slowed down what remained of Cycle~23, and led to the unusually long and quiet minimum of 2008--9.
The concomitant sharp drop in dipole moment adds credence to that claim.
Note that there is no similar feature at the similar post-maximum years in any of the other three cycles, despite all having relatively large post-maximum X-flares. 
The second interesting feature in Figure~\ref{fig:wt}
is marked by the dotted lines, which represent the ``pre-Terminators''
as defined by \cite{Chapman2020}, and which, as mentioned above, correspond to a sharp drop (to virtually zero) in the number of X-flares and strong (high-$aa$ index) geomagnetic storms.
They determined the Cycle~24 pre-terminator to be in August 2016 (and we shall return to studying that most recent, and also most observed, event in greater detail later).
We shall also return to studying flares later,
but in the meantime we observe that the dashed Terminators and dotted pre-Terminators do bracket the periods of higher quadrupole moment and higher variability in the quadrupole and octupole moments.

% {\tt ``Antlion.''}

\begin{figure}[htp]
\centering
\includegraphics[width=\linewidth]{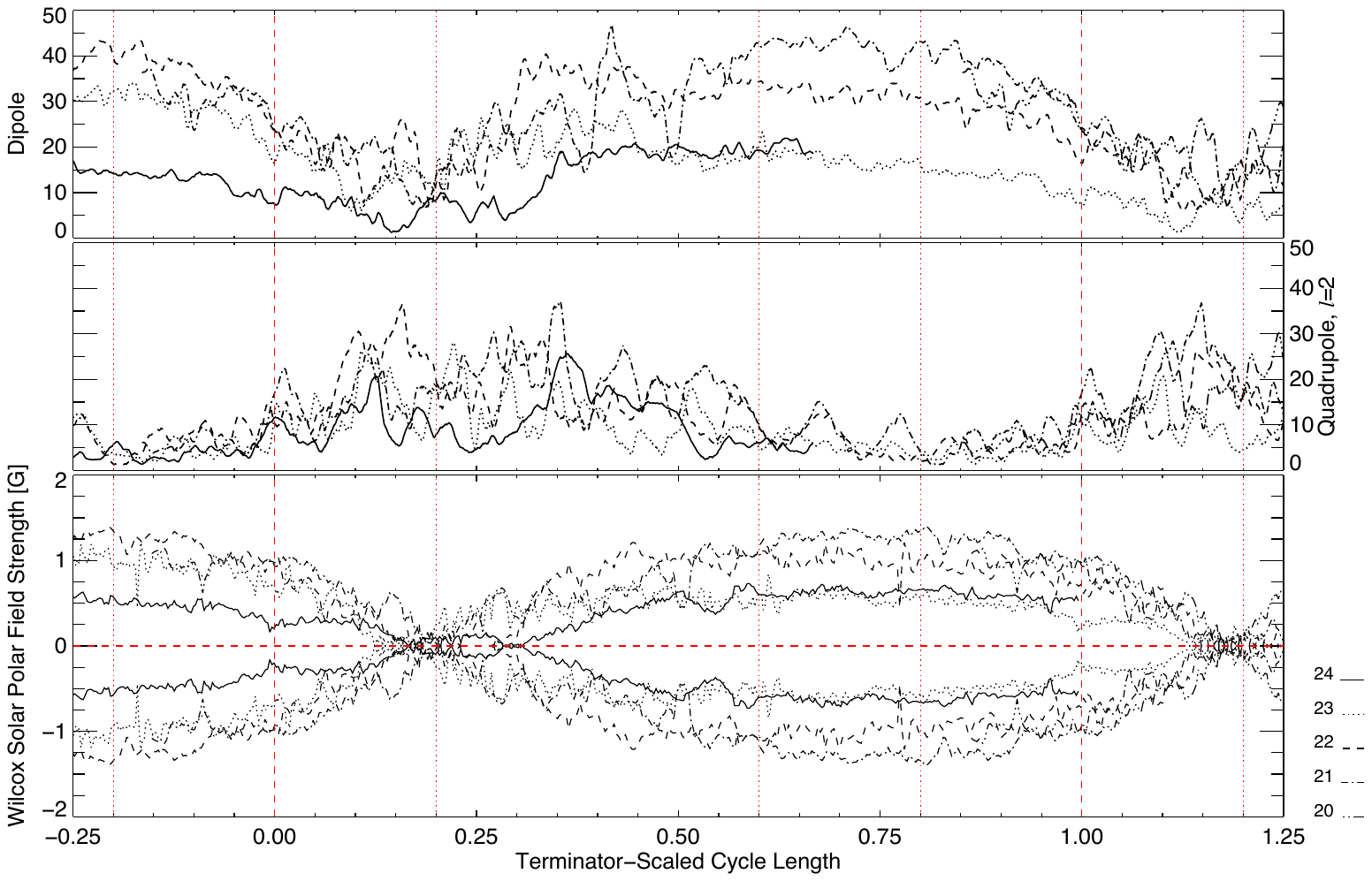}
\caption{(top, middle) Modified Superposed Epoch Analysis version of the Wilcox magnetic field expansion data of Figure~\protect\ref{fig:wt}. (bottom) Modified Superposed Epoch Analysis of the solar polar field as measured by Wilcox from 1976-present. Plotted is the average of the North and South polar apertures, $(N-S)/2$, and again with flipped sign.
The dotted line at $x=0.2$ corresponds to the polar field reversal (canonical max), consistent across all 5 cycles; the dotted line at $x=0.8$ corresponds approximately to canonical solar minimum; and the dotted line at $x=0.6$ corresponds to the pre-Terminator.}
\label{fig:wm}
\end{figure}

Figure~\ref{fig:wm} shows the role of the pre-terminator in bracketing the higher moments more clearly.
The top two panels 
show the same dipole and quadrupole moment data as Figure~\ref{fig:wt}, but
expressed in a ``Modified Superposed Epoch'' analysis
\citep{1913RSPTA.212...75C,2021ESS....801223L}.

Expressing cycle progression as a fraction of their length requires the terminator of cycle 24 to be hard-wired. 
It is set to be 
{July~1, 2021, 
based on the F10.7 measurements discussed in Section~\ref{sec:cbp} and the observed X-flare from AR~12838} 
(this date was just on 
on the outer uncertainty limit of
our extrapolation of the equatorial progression of EUV BPs \citep{10.3389/fspas.2017.00004,LeamonEA2020}).
If Cycle~24 doesn't terminate until later, then the corresponding traces in Figure~\ref{fig:wm} would be compressed leftwards.
As an exercise, we force the terminator date out in time, so as to compress the Cycle~24 trace leftwards in the Figure. 
In order to get the last polar field reversal (at $x \simeq 0.3$) to align with the other cycles' reversal at $x \simeq 0.2$, the terminator date would have to be as late as January 2025, when most, if not all, 
reasonable
solar cycle predictions have solar max occurring at or before that date. 
Further,
an inspection of 
the F10.7 record and new sunspots appearing on the disk 
does suggest that 
{mid~2021 is not a terrible} 
assumption at the time of writing.

Each of the five (partial) cycles 20--24 has its own trace in Figure~\ref{fig:wm}, and the first half of the plot repeats so that the terminators, 
the main objects of focus, 
are not at the edges of the plot.
The dotted line at $x=0.8$ approximately represents canonical solar minimum, and is included merely for reference.
The terminators, represented by the dashed lines at $x=0$, and the pre-terminators, represented by the dotted lines at $x=0.6$, roughly bracket the quadrupole moment.
The more striking result is in the top panel, where the dipole moment shows a clear, coherent, pattern with the minimum ({\em i.e., sunspot\/} maximum) occurring near $x=0.2$, and the maximum dipole occurring approximately at $x=0.6$, {\em i.e.,} the pre-terminator.
This is more clearly seen in the bottom panel of Figure~\ref{fig:wm}.
Rather than the Dipole field strength, here we plot the polar field strength, 
as measured by the Wilcox Solar Observatory
\citep{1978SoPh...58..225S}.
In this Modified Superposed Epoch analysis, 
the polar field reversal happens almost rigidly at 
$x=0.20$ 
($0.198 \pm 0.013$ for the 4 cycles 21--24).
The symmetry of 
one-, three-, and four-fifths of a cycle
is compelling; for reference, the extra 0.01 does appear significant; 0.01 cycle corresponds to about 40--45 days, depending on the length of the cycle, or about 1.5 Carrington rotations.
Canonically, the polar field at minimum $x \sim 0.8$ is used as a precursor predictor method for the upcoming cycle strength; however, we see here that there is little change in the field strength between the pre-terminator at $x=0.6$ and minimum at $x \sim 0.8$.
We must thus question the validity of the polar fields as a predictor of the next solar cycle strength.
However, even
though the superposed epoch analysis of Figure~\ref{fig:wm} is keyed 
{\em on the Terminators},
the polar field reversal happens at the same phase of the cycle;
the {\em timing\/} of
the system is almost predictable.

\subsection{Flare Production}

We have already noted that the first X-flare of a cycle happens close to the Terminator, and \cite{Chapman2020} noted that the pre-Terminator on the declining phase of the solar cycle ushered in the quiet interval spanning solar minimum.
We shall now investigate the production of X-flares throughout the solar activity cycle.
The top panel of Figure~\ref{fig:money} shows the occurrence of all 452 X-flares in the GOES catalogue since 1976, plotted as diamond symbols over the trace of the (27-day smoothed) F10.7 radio flux.

With flares often occurring in close succession and the overlap of plot symbols on a 44-year plot, it might not immediately be obvious that there are 452 points in the top panel of Figure~\ref{fig:money}. 
In fact, 
the vast majority, 96\%, of all X-flares happen between the Terminator and pre-Terminator; 
there are only 16 ``contrary'' X-flares in the entire record.
Four of those are from AR~12673 in September 2017, and the F10.7 flux surged significantly at that epoch (when that AR was present), and exceeded the 
F10.7 $= 90$sfu threshold that is a reasonably good approximation to both the terminators and pre-terminators.
Repeating this analysis for M-class flares, ``only'' 93\%\ happen between the Terminator and pre-Terminator (only 403 out of 5967 fall outside the range). 
The second phenomenon not immediately visible in the top panel of Figure~\ref{fig:money} is that 78\%\ of all X-flares happen when the daily F10.7 value exceeds the 365-day trailing F10.7 average.
% This probably is an oversimplification/ undercount.
This is not a surprise, as the source of X-flares are large complex active regions that generate significant amounts of F10.7 emission due to their size and complexity.
Nevertheless, a refinement of the annual-average simplification could be developed into a useful discriminator of a given AR being likely to have an X-flare. 
% {\color{blue}
To further emphasise the ``pre-requisite'' nature of F10.7 for X-flares, we note that only 8 ($<2$\%) X-flares occurred on a day with the F10.7 reading below 90\,sfu, and only 13 ($<3$\%) occurred in a Carrington rotation with the rotation average F10.7  below 90\,sfu.  
% }
%
Table~\ref{tab:atitle} quantifies the number of X-flares per cycle and the both of these probabilistic observations.

%% Cycle | T-date | Surge | Not Surge | pre-T date | N following pre-T |  

\begin{table}[t]
    \centering
    \begin{tabular}{c|c|ccc|cc}
      \hline
         Cycle & Terminator   & Total  & F10.7 Surge  & Fraction & Pre-T & Post-pre-T  \\
         {}    & Date ($N-1$) & X-Flares & X-Flares       & {}  & Date  & X-Flares  \\
      \hline
         20 & {}       & {}  & {}  & {} & 1973 Apr & 3 \\
         21 & 1978 Jan & 166 & 128 & 77\% & 1984 Jul & 4 \\
         22 & 1988 Jun & 110 & 78  & 71\% & 1993 Aug & 1 \\
         23 & 1997 Aug & 125 & 103 & 82\% & 2005 Sep & 4 \\
         24 & 2011 Feb & 48  & 41  & 85\% & 2016 Aug & 4 \\
      \hline
    \end{tabular}
    \caption{Distribution of all 452 X-flares in the GOES catalogue, relative to the Terminators, pre-terminators and the relative levels in F10.7 radio flux}
    \label{tab:atitle}
\end{table}

The other three panels of Figure~\ref{fig:money}
provide context for 
the abrupt onset of activity at the Terminator and 
the abrupt turn-off of activity at the pre-Terminator.
%% 2
The second panel shows the off-limb (integrated 1.15--$1.25 R_\Sun$) coronal Green line (Fe~{\sc xiv}, 5303\AA) intensity as a function of latitude and time. 
Daily observations are smoothed with a 150-day rolling window.
The {\em overlapping\/} activity bands of the extended solar cycle as discussed above are clearly visible.
At the terminator, the old cycle bands, well, terminate.
The next cycle bands are now at $\sim$35\degree. 
At the highest latitudes, we see the ``rush to the poles'' \citep{Alt97} between the terminator and polar field reversal/ solar maximum.
After polar field reversal, the {\em next\/} cycle's activity band starts progressing equator-ward, starting from $\sim$55\degree. 
%% 4
% Green line Fe XIV, 1.8e6 K
% 195/193 Fe XII, 1.5e6 K
The fourth panel uses the same off-limb sampling process, applied to space-borne broadband EUV imagers from SOHO/EIT (195\AA{}, 1996--2010) and SDO/AIA (193\AA{}, 2010--) to extend the Green line observations to the present.
Not unsurprisingly, the same features are seen between the Fe~{\sc xiv} Green line and Fe~{\sc xii} EUV emission.
%% 3
The remaining panel of Figure~\ref{fig:money} reprises the Solar Polar Field strength of Figures~\ref{fig:wt} and~\ref{fig:wm}.
Through all 4 panels we trace vertical lines for the key landmarks of the cycle: dashed for terminators, thick dashed for polar field reversals, and dot-dashed for the pre-terminators.
As in Figure~\ref{fig:wt}, the Halloween storms of 2003 are marked by the single dotted vertical line.
The pre-terminators occur when F10.7 $\simeq 90$sfu, and at the same time the polar field strength stops increasing and plateaus.

Figure~\ref{fig:msea} re-expresses the data of Figure~\ref{fig:money} in the form of the modified superposed epoch analysis, akin to Figures~\ref{fig:wm} to~\ref{fig:wt}.
We omit the EIT/AIA EUV data as it barely covers two solar activity cycles.
Instead it is replaced by the Calcium~{\sc ii}~K record of cycles 14--20 ($\sim$1913--1977) from the Kodaikanal Observatory, 
as compiled by \citet{2019ApJ...874L...4C}.

Combined, Figures~\ref{fig:money} and~\ref{fig:msea} 
brings everything into focus, and explain the title of this paper.
At the terminator, there is no more old cycle flux/ activity bands at the equator. 
The new cycle bands at mid-latitudes activate, and we observe an increase in coronal emission and flaring. 
At high latitudes we observe the rush to the poles 
(following that of the polar crown filaments and the 
Green Line coronal structures
One-fifth of a cycle later, we observe the polar field reversal.
Two-fifths of a cycle later again, 
at the pre-terminator,
the new (\ie\ next) cycle bands just starts moving equatorward at 55\degree{} 
(as seen in the EIT, Green Line and Calcium~{\sc ii}~K intensity panels), 
and the current cycle bands are at about 15\degree\ latitude.
Independent of the actual mechanism of ``quenching''
the presence of four (two, {\em of opposite polarity}, in each hemisphere) bands on the disk leads to high-energy flaring activity
falling off a cliff.

\subsection{Spectral Irradiance}

\citet{2021ESS....801223L} demonstrated that there was a significant surge in EUV spectral irradiance across the cycle 23 terminator, on a timescale of less than a solar rotation, and possibly as little as a week.
Their results from SDO/EVE, considering the irradiances of the various AIA EUV lines, 
is reproduced as the left-hand panel of Figure~\ref{fig:eve}.
Due to the failure of the SDO/EVE MEGS-A sensor in May~2014, a direct comparison to the 2016 pre-terminator is not possible.
However, a partial comparison can be made using EVE's single-band diodes for 304\AA, 171\AA, 335\AA, and Lyman-$\alpha$ (1216\AA{}), and is shown in the right-hand panel of Figure~\ref{fig:eve}.

The strongest response is seen in  
Fe~{\sc xvi} 335\AA, 
which shows a factor of $\sim$2 change over the course of one solar rotation at both the Terminator {\em and\/} pre-terminator.
Lyman-$\alpha$ shows $\sim$6\%\ change at both times too; smaller change amplified by sheer number of photons---Lyman-$\alpha$ is the most intense line of the solar spectrum at 
0.007 $\mbox{W m}^{-2} \mbox{ nm}^{-1}$ at minimum, rising to 
0.009 $\mbox{W m}^{-2} \mbox{ nm}^{-1}$ at the peak of C24 or similar weak cycles, or 
0.01 $\mbox{W m}^{-2} \mbox{ nm}^{-1}$ at the peak of relatively strong cycles \citep{2015A&A...581A..26L,2018GeoRL..45.2138K}.
% [REFs Kretzschmar, Snow, and Curdt (2018), Lemaire et al. (2015)].
%
Thus a 6\%\ change at the terminator (and pre-terminator) represents approximately one quarter of the entire min-to-max cycle variation, and thus has
significant ionospheric and upper atmosphere impact
\citep{1994JGR....9920665C,2000GeoRL..27.2425L,2021ESS....801223L}. 
% (Chandra and McPeters, 1994, Lean 2000) 

We reiterate that this step change in radiative output occurs independent of wavelength, and is co-temporal with the drop-off in high energy flares.

\subsection{Sunspot Complexity}
\label{sec:kurt}

EUV Photons emitted from hot coronal plasma and energetic flares both require significant energy to be supplied to the corona from below.
Why does the energy available for coronal energetics so sharply and dramatically reduced?
At the pre-terminator, the current cycle band is at $\sim$15--20\degree\ latitude. Two possibilities that explain the decrease in activity related to band location are:
\begin{itemize}
  \item being 30 degrees apart means greater interaction between the two current-cycle bands across the equator, reducing energy available for coronal energetics; or
  \item being at or below some critical latitude means that the north-south shear from differential rotation is too great to support or maintain large complex ARs, reducing energy available for coronal energetics. (We note that the pre-terminator occurs at approximately the Carrington corotation latitude of $\sim$14.5\degree, but this is most likely just a coincidence.) 
\end{itemize}
Either way, as shown in the previous sections, F10.7, occurrence of X-flares, and coronal spectral emission fall off sharply.
We surmise that the turn-off in activity at the pre-terminator is due to a (sudden) decrease in active region complexity.

\cite{2016ApJ...820L..11J}
% Jaeggli and Norton
noted that
``there are no published results showing the variation of the Mount Wilson magnetic classifications as a function of solar cycle,''
and as an attempted rectification,
considered the fraction of complex ($\gamma$ and/or $\delta$) sunspot groups from 1992--2015, 
albeit only on an annually-averaged basis. 
They did find a sharp decrease in the number of the most complex $\beta \gamma \delta$ regions prior to minima in 1995 and 2007, and increase in the fraction of simple $\alpha$ regions, but the direct utility to our study is limited by their annual averaging.

Similarly, 
\cite{2005ApJ...631..628M}
% McAteer, Gallagher, Ireland
considered the fractal dimension of a 50~Gauss threshold contour of all ARs seen by SOHO/MDI from 1996--2004, but focused more on testing a correlation of complexity to flare likelihood.
And, unfortunately, curtailed their study before the Cycle~23 pre-terminator.

In the absence, then, 
of a complete database of active region classifications, 
we can consider the distribution of spot areas from the Greenwich/ USAF record as proxy, which is what is shown in Figure~\ref{fig:cplx}. 
The three panels are the mean, skewness and excess kurtosis of the spot areas on a Carrington rotation-by-rotation basis. 
There are clear Terminator and pre-terminator signals, especially in the higher order moments. 
(Recall excess kurtosis has 3 subtracted from the moment calculation, as 3 is the computed value of a Gaussian distribution.  
Thus excess kurtosis $<0$ means flatter than Gaussian; 
similarly, skewness $>0$ means the distribution is skewed to the right, with a longer tail to the right of the distribution maximum.)

Overlaying the kurtosis panel of Figure~\ref{fig:cplx} for the last half of the record is F10.7. 
The correspondence between kurtosis, and F10.7, simply scaled and shifted, especially from $\sim$1960 (post Cycle-19 maximum) on, is quite remarkable.  
F10.7 really is just an integrated measure of global complexity of the solar magnetic field.

{% \color{blue}
\subsection{The Circle of Fifths}

We complete our description of the Hale Cycle with one further 
analysis of the F10.7 radio flux.
The top panel of Figure~\ref{fig:unit} again shows the complete record of F10.7, with the Terminators and pre-terminators marked as vertical dashed red lines. Here we add markers of 0.2~cycle and 0.4~cycle as vertical dotted lines. In all cycles, and markedly so in Cycles~20 and~22, there is is a distinct drop-off from its maximum (solar maximum) value at around 0.4~cycles.
To investigate this further, we compute a unit cycle, following the methods of \citet{2021ESS....801223L}.

To create the bottom panel of Figure~\ref{fig:unit}, we take the almost 7 cycles of F10.7 data from the top panel, subtract off the minimum value of 65.0~sfu, and scale each cycle to the maximum of its (13-month smoothed) record.
We then scale time as a fraction of Terminator-Terminator time,
interpolated to 100 points, 
and plot the mean and standard deviation for each fractional time (thick black line and red hatched envelope).
There 
is consistently: 
\begin{enumerate}
    \item 
a rapid surge in F10.7 (through 90~sfu, almost by definition) at the Terminator; 
    \item 
a dip around polar field reversal at 0.2 cycles; 
    \item 
a sharp drop at 0.4 cycles 
when elements of the {\em next\/} cycle appear at high latitudes;
    \item 
the onset of ``quiet'' conditions at 0.6 cycles 
as discussed before, 
and
with approximately the same F10.7 output as at the Terminator; 
    \item 
and finally 
0.8 cycles is approximately canonical minimum.

\end{enumerate}

%% The Law of Fifths. The Circle of Fifths. All explained by landmarks of the 22-year Hale Magnetic Cycle.

The bottom panel of Figure~\ref{fig:unit} further shows that there is also a clear change in the standard deviation (thickness of red envelope), at the Terminator, 0.4, and 0.6 cycles. 
Regardless of the strength of the cycle or whatever short term activity surges are superimposed on it, the nature of (the complexity of) sunspots changes at these landmarks of the 22-year Hale Magnetic Cycle.

% For 
Finally, for
an approximately 11-year cycle, 
we note
there are approximately 150 Carrington rotations per cycle, so each fifth/ phase of the cycle takes approximately 30 rotations.

}

\section{The Solar Cycle Clock}

We are now in a position to put together a coherent explanation of all solar activity measures described above, predicated on the existence of multiple activity bands, per \cite{Mac14}.

Let's start our story at the terminator. 
With the cancellation of old-cycle polarity flux at the equator,
a tachocline tsunami \citep{2019NatSR...9.2035D} triggers the rise (literally and metaphorically) of the new cycle band, which is now at 30--35\degree\ latitude. Larger sunspots form these mid-latitudes (with some spread due to the rise of fluxropes through the turbulent convection zone). 
Changes in buoyancy also give an increase in active region longitudes from $\sim$1 to $\sim$5 (Rossby modes). 

The presence of new active regions 
reconfigures the large-scale coronal magnetic field, and thus heliospheric field, warping the heliospheric current sheet.
The heliospheric field changes causes a sharp ($\sim$4\%) drop in GCR flux 
\citep{1954JGR....59..525F}; potential implications for Earth's atmosphere are discussed by \citet{2021ESS....801223L}.
The surge in coronal heating from larger active regions gives rise to energetic changes such as F10.7 and EUV 
which also have terrestrial atmospheric (weather) implications: 
the North Atlantic Oscillation changes sign from negative to positive  due to EUV-driven stratospheric heating and photochemistry changes 
\citep{1995Sci...269..676H}.
A hotter corona also drives changes in Alpha composition as observed in in situ solar wind data. 
The first X-flare of the new cycle happens within a solar rotation of the Terminator. 

Back in the corona and chromosphere, the ``rush to the poles'' is seen in the coronal green line \citep{Alt97} and polar crown filament data
\citep{2020arXiv201006048M}. 
The poleward progression of the PCF---the highest latitude magnetic neutral line---starts the polar field reversal process.

The rise phase of the solar cycle occurs, with increasing sunspot number, sunspot complexity, and weakening polar fields, and continues until the (opposite polarity) bands of cycle $n+1$ first emerge at 55\degree\ at (well, defining) maximum. 
The reversal of the polar fields is not rigidly tied to the maximum sunspot number, but does (rigidly) occur at solar cycle phase 0.2. 
{% \color{blue}
There does appear to be dip in F10.7 radio flux at the time of polar field reversal. 
The obvious explanation for this is that the two hemispheres do
not peak at the same time 
\citep[certainly true for the majority of cycles; see, {\em e.g.},][]{2021A&A...652A..56V}, but also 
that the individual hemispheres show separate surges of activity that lead to a double peak 
\citep{2015NatCo...6.6491M, 2017NatAs...1E..86M}.

At phase 0.4 of the cycle, the polar fields have returned;  
0.4 cycle also corresponds to
the re-formation of the polar crown filament at $\sim$55\degree, the disappearance of all higher-latitude filaments \citep{2020arXiv201006048M}.
Recall also the Green Line (panel~b) and, in particular, the Calcium~{\sc ii}~K panel (bottom panel) of Figure~\ref{fig:msea}: where there is almost no Ca~{\sc ii}~K emission above $\sim$55\degree\ for the whole interval 0--0.4 cycles and 0.4 cycle marks the strengthening of Green Line emission at 55\degree, and the start of the equatorward progression of such features.
There is a noticeable change in F10.7 flux (but less of a change in sunspot number) as the interaction of the now two activity bands below 55\degree\ constrains the global active region complexity.
}

On the downslope of the cycle, both old and new bands progress equatorwards, and interactions between the two give rise to more and greater flares, eruptions and storms post-sunspot maximum, 
{% \color{blue}
even as F10.7 and the sunspot number are sharply reduced.}

The slower this equatorward progression (\ie\ the longer the cycle) the more interactions there are between bands, which is why the empirical cycle length--next cycle strength relationship exists 
{% \color{blue}
\citep{2015LRSP...12....4H,McIntoshEA20b}}. 

At the pre-terminator, 
at cycle phase 0.6, 
the current cycle band is now at $\sim$15-20\degree{} latitude. As pointed out in the previous subsection, two possibilities that explain the marked decrease in activity here are
{% \color{blue} 
the bands are close enough across the equator to interact, or the north-south shear from differential rotation is too great to support or maintain large complex ARs.
Either way the energy available for coronal energetics is reduced, and} F10.7, alpha composition, and flare production fall off a cliff. There are (almost) no more X-flares. 

Smaller active regions do still form, which is why the smoothed SSN curve has the quasi-Maxwellian form \citep{2015LRSP...12....4H} 
that it does, 
% {\color{blue} 
and the relationship between sunspot number and  sunspot activity belt central latitude has the quasi-linear form \citep{Waldmeier55.book,2016A&A...591A..46C} 
that it does,
% }
even though activity does not follow (no longer does) SSN. 
Put another way, more bluntly, 
SSN could be considered a ``MacGuffin.''
% SSN IS AN EFFIN' MAGUFFIN!

We suspect that the increase in mid latitude coronal holes from the next cycle band grow at the pre-Terminator too, giving the $aa/ A_p$  predictability of \cite{Feynman82}, but we reserve study of coronal holes to a subsequent 
(student-led) 
paper.

Between the pre-terminator and solar minimum, the old cycle's bands weaken due to cancellation with their opposite-polarity counterparts across the equator, and the new cycle bands continue to grow in strength in the mid-latitudes;
as discussed in 
\cite{2020arXiv201006048M},
% STEVE!
and the Introduction above;
solar minimum occurs when all 4 bands have approximately equal strength, at cycle phase 0.8. 

As the new cycle bands continue to gain dominance, small new-polarity regions begin to rise and appear, maybe some even producing mid-level flares, but it is not until the Terminator do large regions and large flares occur.

The cycle repeats.

Finally, we note that the above describes a single activity cycle, approximately 11 years long.
\cite{Chapman2020} 
also expressed the F10.7 and $aa$ index activity as a clock with an average 11 year cycle;
\citet{Chapman2021} 
extend the concept to the 
difference between 11- and 22-year cycles.
The two different (solar polar) polarities that comprise the Hale Cycle  affect the inflowing galactic cosmic rays
\cite[e.g.,][]{1977ApJ...213..861J,1981ApJ...243.1115J,2014SoPh..289..407T},  
but also how solar coronal activity translates into the level of disorder in the solar wind.  
The data shown in Figures~\ref{fig:wt}--\ref{fig:eve}
barely span two full Hale cycles, so we do not attempt to generate ``clock'' figures with 
a rigid terminator-terminator phase as the ``hour hand,''
and instead refer the interested reader to the companion papers
\citep{Chapman2020,Chapman2021}.
% Chapman et al.\ papers.

\section{Summary and Discussion}
\label{sec:summ} 

Through combined observations of the Sun's radiative, particulate and ejecta output, 
we have demonstrated that there is a clear {\em Solar Clock\/} defined by the Terminators of each 22-year Hale Magnetic Cycle.
Landmarks of this cycle, such as 
the first and last X-class flare, 
F10.7 exceeding and falling below a certain threshold, 
the poleward motion of the highest latitude filaments, and
the reversal of the polar field 
all occur at fixed points in this 
(unit) 
cycle.

The repeating global-scale magnetic pattern that first appears at $\sim$55\degree\ latitude, with one branch rushing to the poles over 2--3 years and the second progressing equatorward over 19--20 years, is key to understanding the Hale Magnetic Cycle. Other than the rush-to-the-poles time, there are four oppositely signed magnetic bands between 55\degree\ and the equator, and their interactions drive solar activity.

\begin{figure}[p]
\centering
\includegraphics[width=0.95\linewidth]{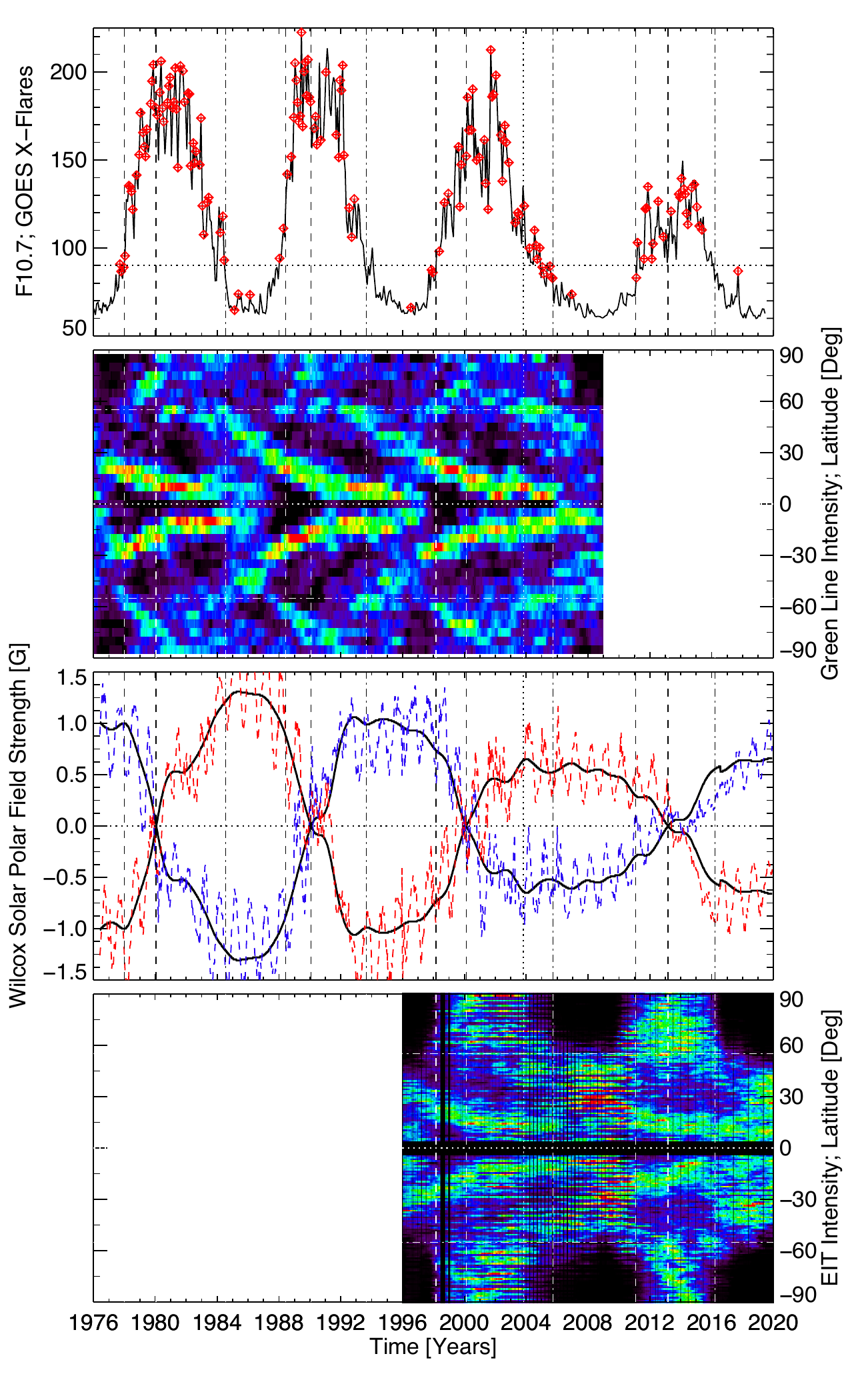}
\caption{Showing the importance of the pre-Terminator across all aspects of solar output.
(a)~F10.7 solar radio flux, overplotted with the time of GOES X-class flares; 
(b)~coronal Green Line emission as a function of latitude; 
(c)~Wilcox Polar field strength, as in Figure~\protect\ref{fig:wm} (red--south, blue--north, and the smoothed average in black);
(d)~EIT corona, akin to panel~(b) and Figure~9 of \protect\cite{Mac14}. Green Line and F10.7 data can be extended back several more decades; we limit the time axis to 1976--2020 for GOES flare and Wilcox Observatory data. Throughout all panels, terminators are marked by dashed lines, the pre-Terminators (F10.7 = 90~sfu) by dot-dashed lines; solar maxima (as determined by polar field reversals in panel (c)) are also marked by dashed lines.
The dotted vertical line corresponds to the Halloween Storms of 2003.
}
\label{fig:money}
\end{figure}

\begin{figure}[p]
\centering
\includegraphics[width=0.95\linewidth]{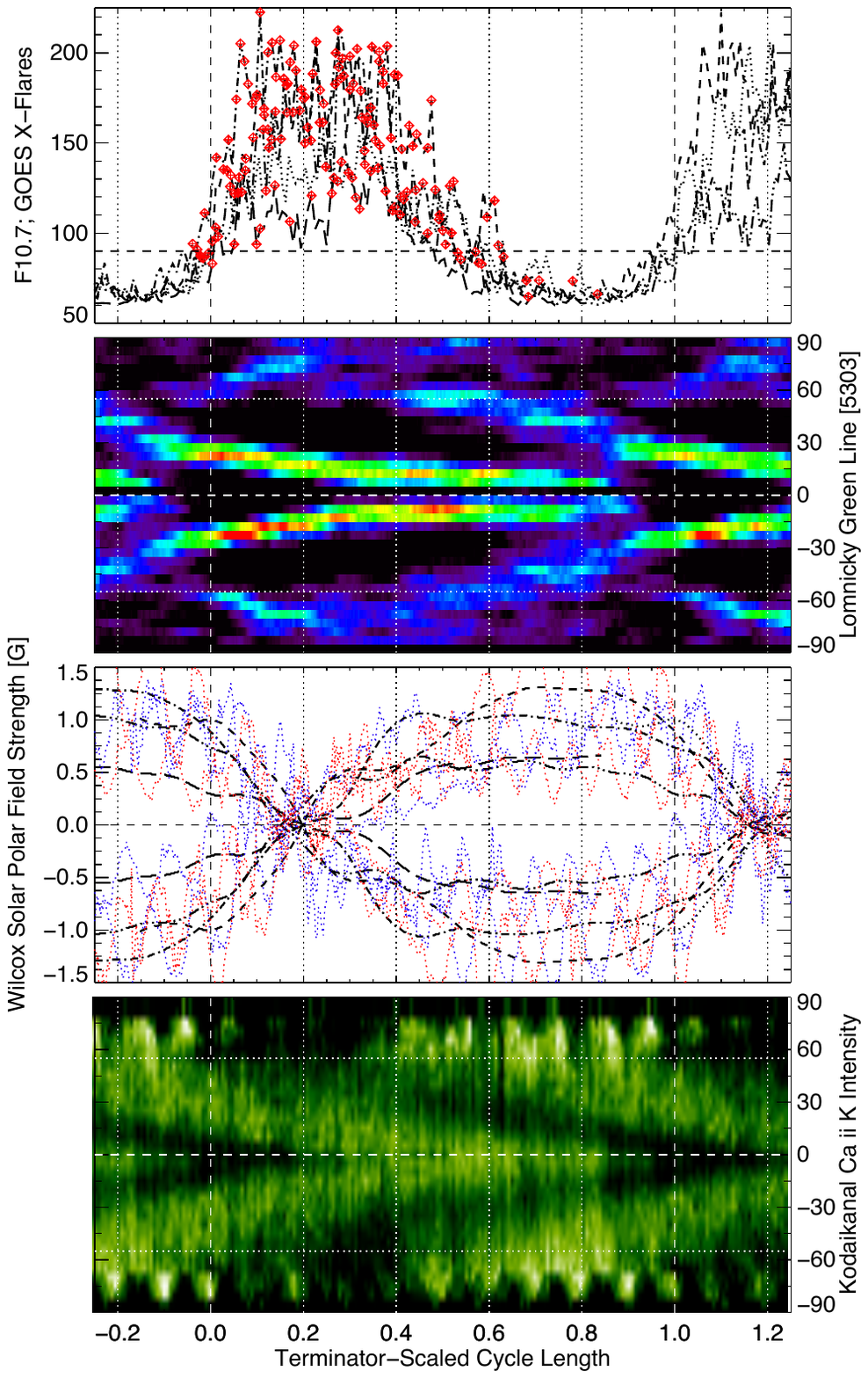}
\caption{Modified superposed epoch analysis version of Figure~\protect\ref{fig:money}.
The EIT panel is replaced by the Calcium~{\sc ii}~K record of cycles 14--20 ($\sim$1913--1977) from the Kodaikanal Observatory.}
\label{fig:msea}
\end{figure}

\begin{figure}[p]
\centering
\includegraphics[width=0.475\textwidth]{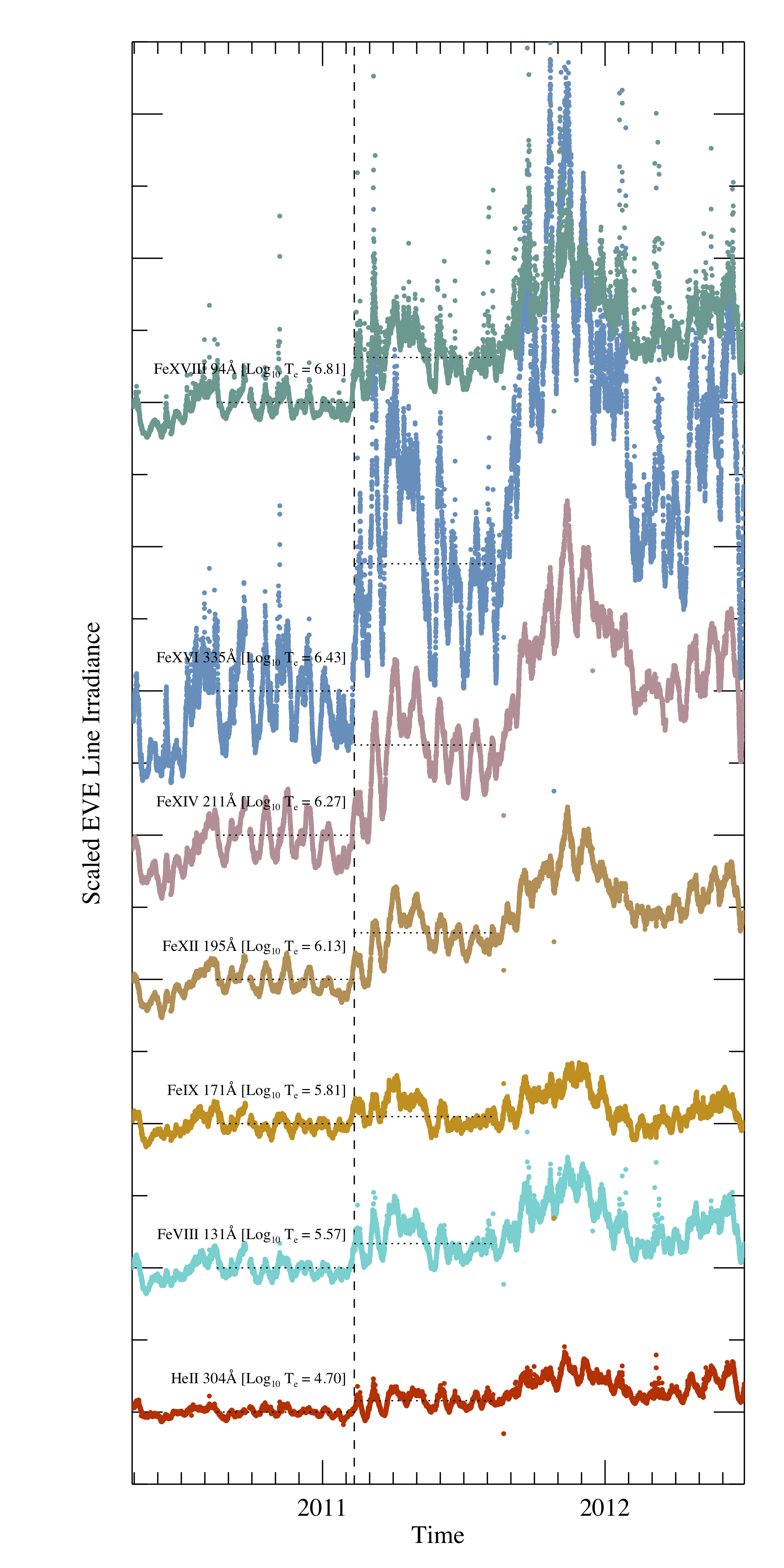}
\includegraphics[width=0.475\textwidth]{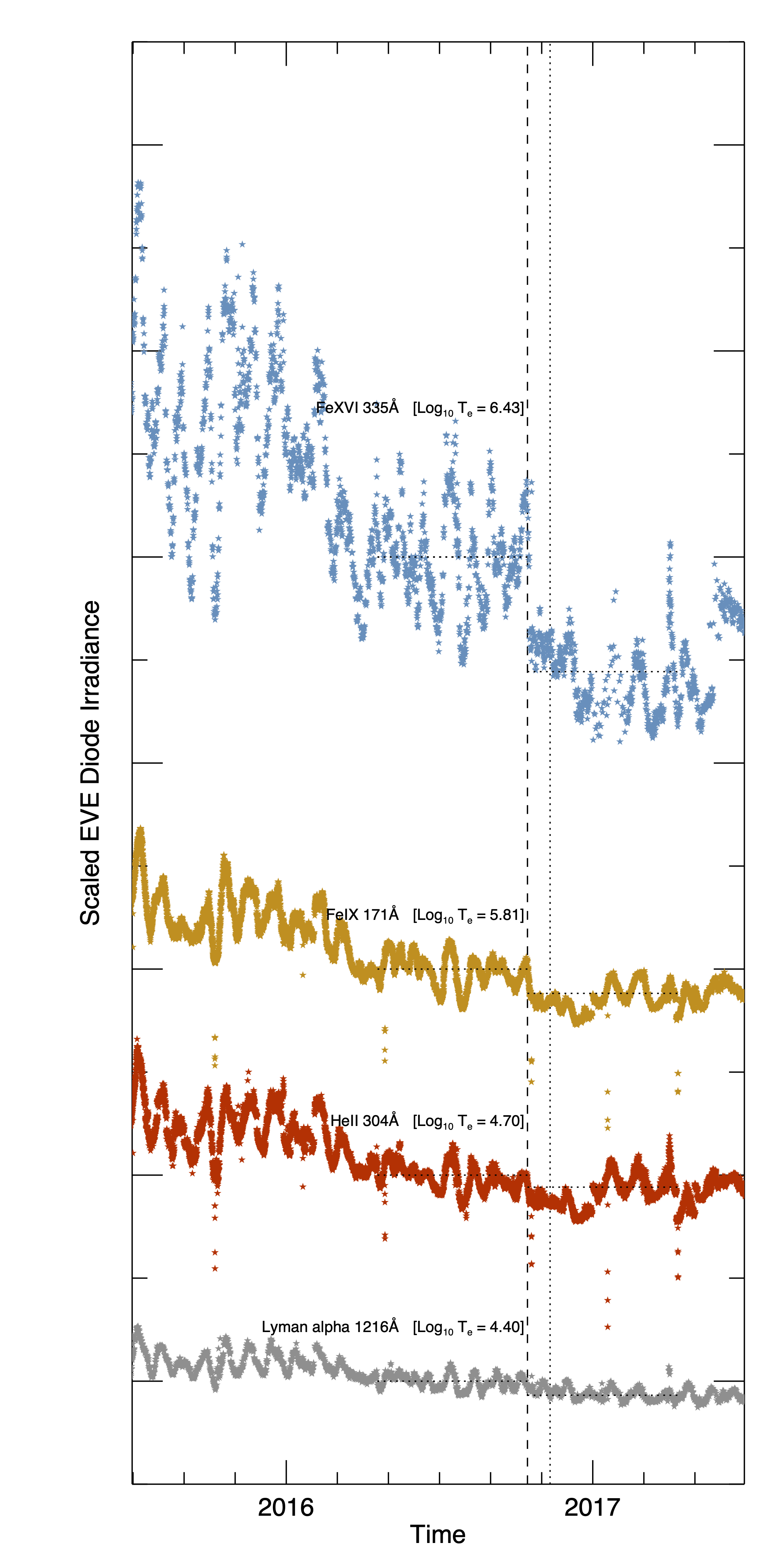}
\caption{The evolution of constituents of the the Sun's spectral irradiance, as measured by the SDO/EVE diodes, across the {\em sharp\/} transition at 2016 pre-terminator. From bottom to top, the lines increase in their formation temperature. Each successively hotter line is scaled to the time interval 180–60 days prior to the pre-terminator, and offset on the $y$-axis by unity, except for the hottest line---Fe~{\sc xvi} 335\AA---which is offset by two to show the huge change in its emission. The pre-terminator is indicated by the dashed vertical line; the dotted vertical line represents one Carrington Rotation later. Drop-offs: Lyman = 0.93; 304\AA\ = 0.94; 171\AA\ = 0.88; 335\AA\ = 0.51. (The left-hand panel is reproduced from Figure~2 of \protect\cite{2021ESS....801223L}, and shows a $\sim$180\%\ increase in Fe~{\sc xvi} 335\AA\ emission across the terminator.)}
\label{fig:eve}
\end{figure}

\begin{figure}[p]
\centering
\includegraphics[width=\textwidth]{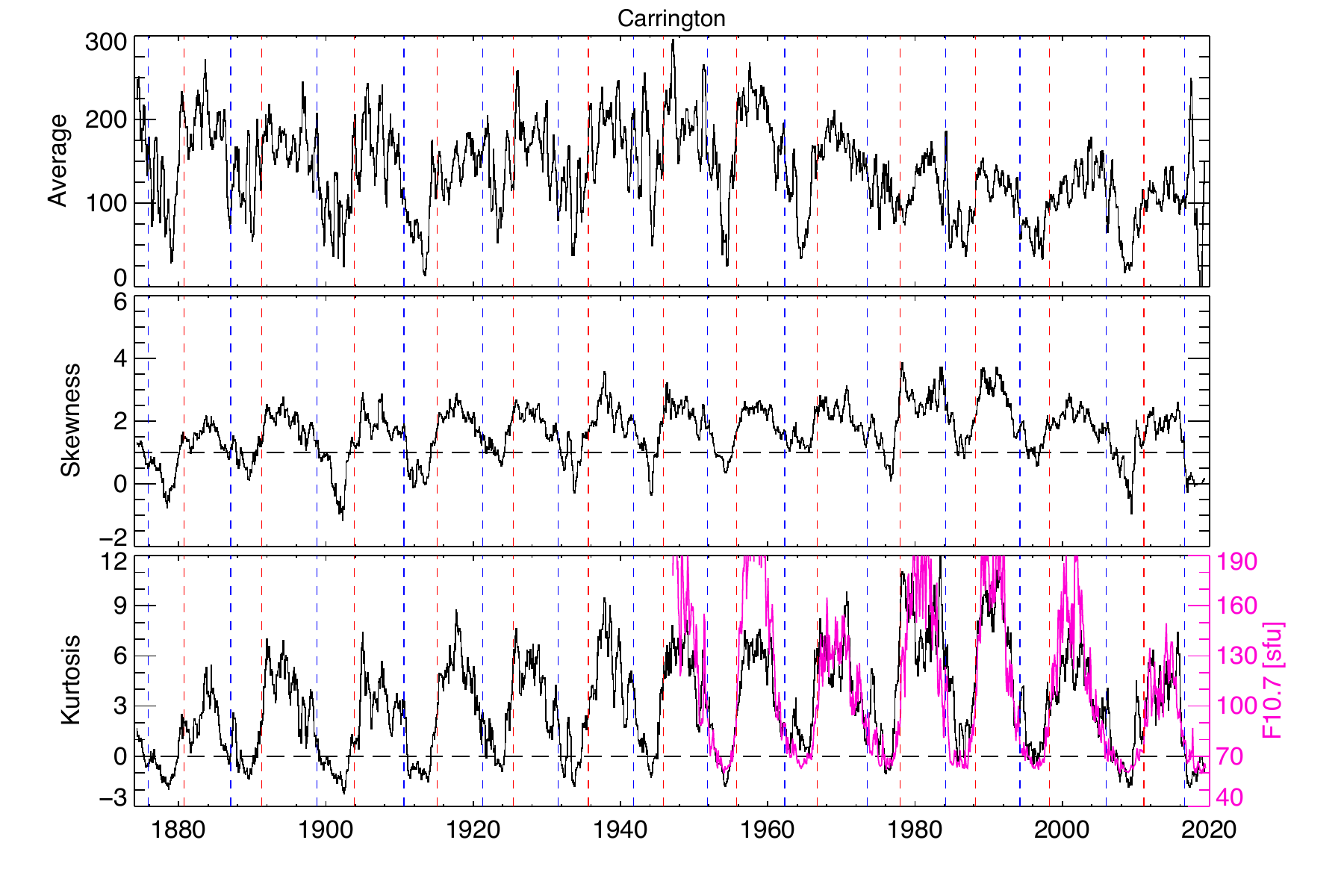}
\caption{The Complexity of sunspots changes at Terminators and Pre-Terminators. As a proxy of geometric complexity for individual spots, we compute the moments of the distribution of spot area from the Greenwich/ USAF archive. From the top, the three panels are the average, skewness and excess (compared to a Gaussian) kurtosis. Dashed vertical lines indicate the Terminators are in red, Pre-Terminators in blue. The kurtosis is a remarkable proxy for the F10.7 radio flux.}
\label{fig:cplx}
\end{figure}

\begin{figure}[p]
\centering
\includegraphics[height=\textwidth,angle=90]{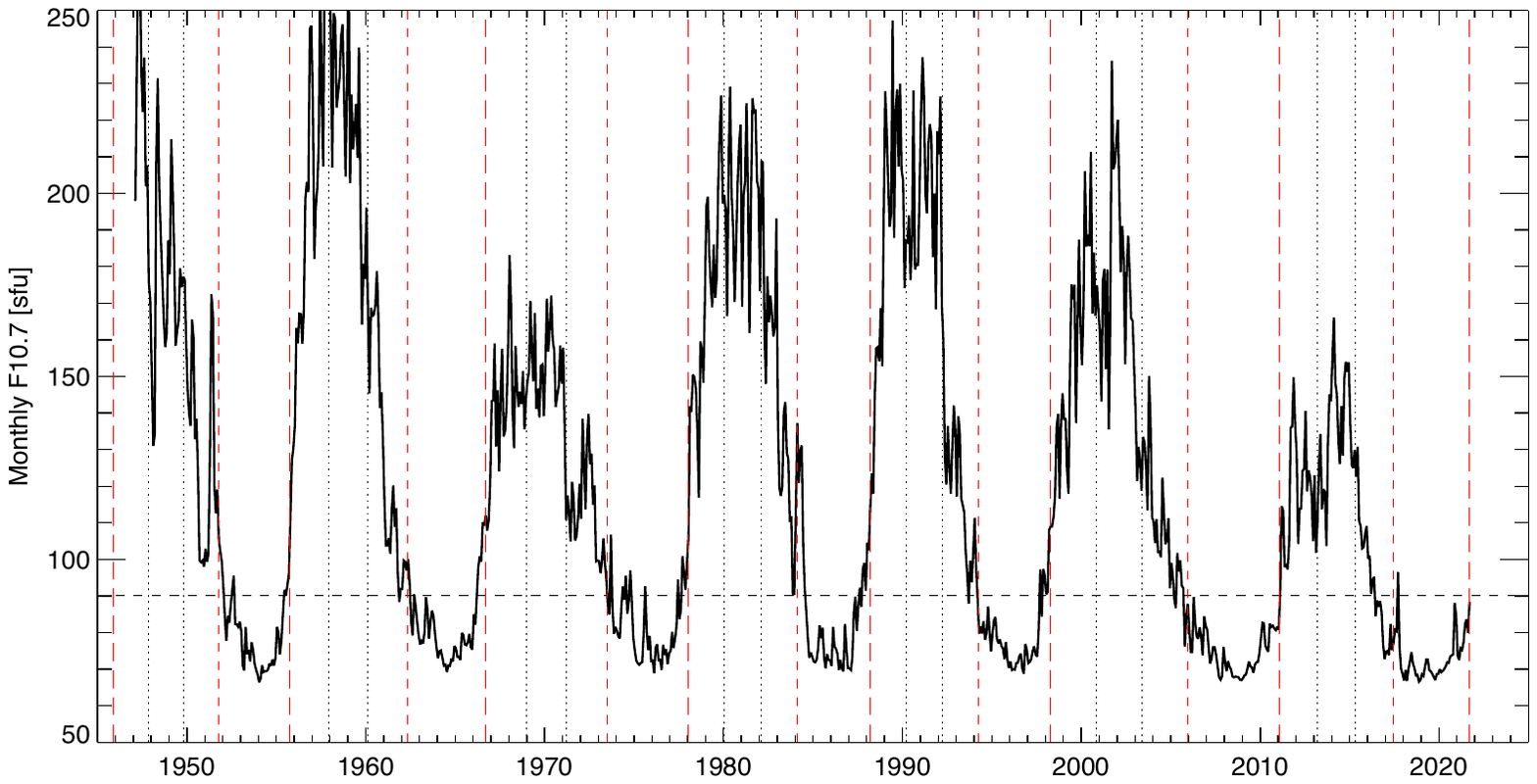}
\includegraphics[width=\textwidth]{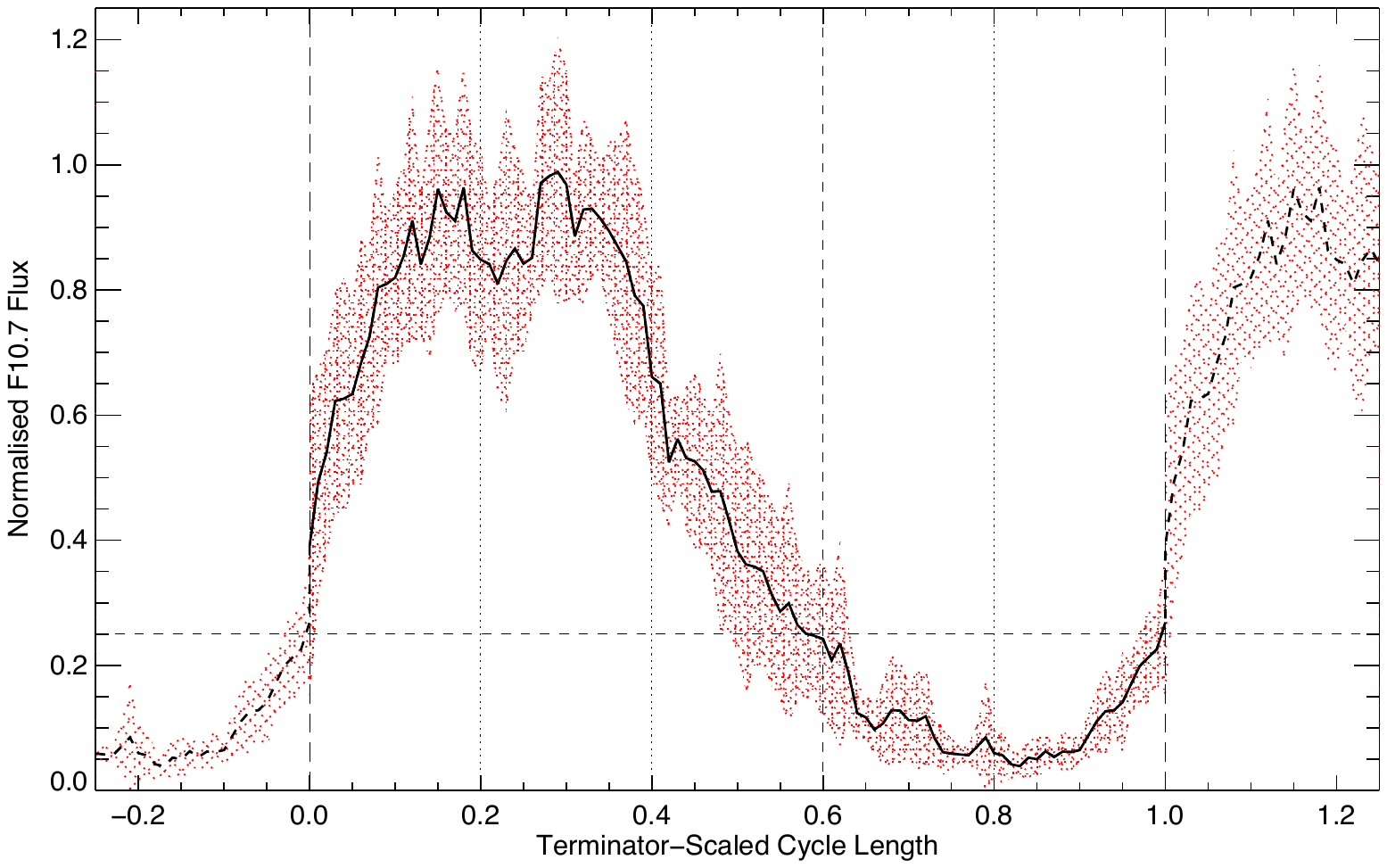}
\caption{(top) The complete F10.7 record, with 0.2, 0.4, and 0.6 of cycle indicated by vertical lines.
(bottom) The terminator-terminator Unit Cycle, scaled to each cycle's maximum radio flux, as computed from the above. The rapid surge in F10.7 at the Terminator, dip around polar field reversal at 0.2 cycles, and sharp drop at 0.4 cycles (when elements of the next cycle appear at high latitudes) are all clearly visible. }
\label{fig:unit}
\end{figure}

Magnetic activity cycles have been observed on many other stars, both solar twins and solar analogues, and stars of other spectral classes.
One would assume that the concept of Terminators 
is also applicable to other stars---for example the sudden turn-on and turn-off of activity in stellar chromospheric Ca~{\sc ii} H-K indices 
\citep[e.g.,][]{2007AJ....133..862H}. 
Similarly, as Figures~\ref{fig:money} (for X-ray flux)--\ref{fig:eve} (for EUV) 
show, reducing solar output to a single scalar time series, 
{\em i.e.}, 
Sun-as-a-star spectral irradiance observations,
have recently been shown to have plasma diagnostic properties
that can
probe the extent of coronal magnetic fields above starspots \citep{2020ApJ...902...36T}.
However, the sparse and non-continuous temporal availability of such observations 
(limited viewing above/ below the celestial equator; 
{\em e.g.}, Hall et al.'s Figures~7 and~8) 
makes for challenging identification of stellar Terminators.

In closing, since 
the flux system that will (imminently) provide cycle~25 activity has been present on the sun and visible in the BP panel of Figure~\ref{fig:bands} since 
just after the cycle 24 solar maximum, 
we can 
{\em already\/}
predict when cycle 25 will end by linear extrapolation of the cycle 25 bands' equatorward progress.
\cite{LeamonEA2020} thus predicted with October~2031 $\pm$ 9~months
(and therefore cycle 25 will be of quite average length), 
which puts the cessation of major solar activity associated with the next 
pre-terminator
in the first half of 2027 (March $\pm$ 6~months).
{% \color{blue} 
These predictions will be refined at the polar field reversal (one-fifth of the way through the terminator-to-terminator cycle)---which we may currently estimate to be the first half of 2024.
We note that according to the recent results of \cite{2021SoPh..296...82O}, 
the likelihood of extreme geomagnetic consequences among 
those events that do occur in $\sim$2024--2027 after the polar field reversal and before the pre-terminator is increased compared to the coming few years (the rise phase of the cycle from now until the polar field reversal).}
For reference,
{\em Solar Orbiter\/}'s 5th Venus Gravity Assist maneuver will occur in December~2026, lifting the peak heliolatitude of its orbit to $\sim$25\degree\ in March~2027, 
which is extremely fortuitous timing to observe the anticipated pre-Terminator changes in 
flux systems at 55\degree.
Two other out-of-ecliptic missions are under consideration
by NASA---the {\em Solar Cruiser\/}
Mission-of-Opportunity is in development (to launch with {\em IMAP}, scheduled for early 2025),
and the 
{\em Solaris\/}
concept is in a Phase-A study for the next MIDEX mission, and if selected will launch in late 2025 and will make its first high-latitude polar pass in mid~2029.

The significant advances these 
coming observations
will make, together with the decade of step-by-step improvements to theory and modeling 
(constrained by said significant observational advances)
they will drive, 
makes it probable that, as it rises in 2021, Cycle 25 will be the last solar activity cycle that is not fully understood.

%% FIN

\acknowledgements
RJL is supported by NASA's Living With a Star Program.
SMC is supported by the National Center for Atmospheric Research, which is a major facility sponsored by the National Science Foundation under Cooperative Agreement No.\ 1852977. 
RJL also appreciates the support of NCAR's HAO Visitor Program, which enabled discussions with SMC, as well as Sandra Chapman and Nicholas Watkins. 
We thank Phil Scherrer and J.~Todd Hoeksema for their assistance with, and discussions on, the Wilcox Solar Observatory data, and Dipankar Bannerjee and Subhamoy Chatterjee for their assistance with the Kodaikanal Observatory data.

%%
%%%% FIGURES 
%%
%%

%%
%%%% BIBLIOGRAPHY 
%%
%%

% \clearpage

\bibliographystyle{spr-mp-sola}
% \bibliography{sample,weenina,scottstevehere}
% \include bbl file for ArXiv submission

\end{article} 

\end{document}